\documentclass[twocolumn,prb,aps,superscriptaddress,showpacs]{revtex4}
\usepackage{graphicx}

\begin{document}
\title {Anisotropic transport for $\nu=2/5$ FQH state at intermediate magnetic field}
\author{Shi-Jie Yang}
\affiliation{Center for Advanced Study, Tsinghua University,
Beijing 100084, China} \affiliation{Institute of Theoretical
Physics, Chinese Academy of Sciences, P.O. Box 2735, Beijing
100080, China}
\author{Yue Yu}
\affiliation{Institute of Theoretical Physics, Chinese Academy of
Sciences, P.O. Box 2735, Beijing 100080, China}
\author{Bang-Fen Zhu}
\affiliation{Center for Advanced Study, Tsinghua University,
Beijing 100084, China}

\begin{abstract}
The $\nu=2/5$ state is spin-unpolarized at weak magnetic field and fully
polarized at strong field. At intermediate field, a plateau of half the
maximal polarization is observed. We study this phenomenon in the frame of
composite fermion theory. Due to the mixing of the composite fermion Landau
levels, the unidirectional charge/spin density wave state of composite
fermions is lower in energy than the Wigner crystal. It means that transport
anisotropy, similar to those for electrons in higher Landau levels at half
fillings, may take place at this fractional quantum Hall state when the
external magnetic field is in an appropriate range. When the magnetic field
is tilted an angle, the easy transport direction is perpendicular to the
direction of the in-plane field. Varying the partial filling factor of
composite fermion Landau level from $0$ to $1$, we find that the energy
minimum occurs in the vicinity of one-half.
\end{abstract}
\maketitle

%\newpage \pagenumbering {arabic}

The Coulomb interaction between electrons in two-dimensions plays
a dominant role when the kinetic energy is quenched into a series
of degenerated Landau levels (LLs), and gives rise to a variety of
unusual phenomena. The most famous example is the fractional
quantum Hall (FQH) effect\cite{book1}, where at certain filling
factors the system condenses into the so-called Laughlin liquid
which acquires gaps for making charge excitations. Another
manifestation of Coulomb interaction is the formation of
unidirectional charge density wave (UCDW) or stripe phases in high
LLs at half fillings\cite {Lilly,Du1,Phillips,Jungwirth}. A
unified description of all the fractional quantum states in the
lowest LLs $\nu=n/(2pn+1)$ was achieved by the composite fermion
(CF) picture, in which each electron carries $2p$ statistical
magnetic quanta to form a composite particle \cite{Jain1,book2}.
At a mean field level, the CFs experience an effective magnetic field $%
B^*=B/(2pn+1)$, in which they fill $n$ CF-LLs, and exhibit integer quantum
Hall effects.

In recent years, the spin polarization of electrons for integral as well as
fractional quantum Hall states have been studied extensively. Many novel and
interesting spin-related phenomena have been revealed. A large amount of
data have been accumulated for various fractional filling states\cite
{Eisenstein,Du2,Chak,Leadley,Khan,Melinte,Freytag} as well as for $\nu=1$%
\cite{Barrett}. In a tilted magnetic field experiment carried out by W. Pan
\textit{et al}\cite{Pan}, the transport becomes highly anisotropic for
even-number filling factors ($\nu=4,6,\cdots$) when the tilted angle exceeds
a large critical value. A new phase called spin density wave was proposed to
explain this phenomenon\cite{Pan,Demler}.

It is now well established that for some filling factors ($\nu=1/m$, $m$ odd
number) the ground state is fully spin polarized for all values of Zeeman
splitting, while for other filling factors (for example, $\nu=2/5, 3/7, etc.$%
) the ground state is fully polarized only for large values of Zeeman
splitting but unpolarized for small (or zero) values of Zeeman energy. One
interesting observation is that at intermediate values of Zeeman energy
there appears a plateau of half the maximal spin polarization for $\nu=2/5$%
\cite{Kukushkin}. The stability of the half-polarized state
implies that the ground state energy of the system as a function
of spin polarization has non-monotonic behavior at half
polarization. The Zeeman energy $E_Z=g\mu_B B_{tot}$ favors
spin-polarization, while the electron-electron interaction favors
a singlet state. However, the origin of the half-polarization
phenomenon remains controversial. According to theory of composite
fermion
with a spin, each CF-LL is split into two subbands. As the $n=0\uparrow$%
-spin band is always fully occupied, it can be treated as a
non-dynamical background. At intermediate $E_Z$, the $n=0\downarrow$-spin CF-LL and the $%
n=1\uparrow$-spin CF-LL can both be partially occupied with filling factors $%
\nu_1$ and $\nu_2$, respectively. $\nu_1+\nu_2=1$. The half-polarized state
corresponds to $\nu_1=\nu_2=1/2$. The composite fermions occupying the
partially filled CF-LLs can be thought as a system consisting of two types
of fermions. V.M. Apalkov \textit{et al}\cite{Apalkov} mapped the
two-component fermion system onto a system of excitons and described the
ground state as a liquid state of excitons with nonzero values of exciton
angular momentum. Their calculation reveals that a downward cusp occurs at
half the maximal spin polarization. On the other hand, G. Murthy\cite{Murthy}
proposed a partially polarized density wave (PPDW) state of CFs based on a
Hartree-Fock theory developed earlier \cite{Shankar}. He compared the ground
state energies of the PPDW state, in which one set of CDW was placed
directly over the other, with the Halperin-(1,1,1) liquid state\cite
{Halperin} and found that the PPDW state is slightly lower in energy than
Halperin-(1,1,1) state\cite{Murthy,Murthy2}.

Activated by the above works, we compare the cohesive energies of
the CDW and the UCDW states for CF-LL filling factor $\nu=2$.
Allowing for different stacking possibilities of the two sets of
interacting CDW or UCDW states consisting of type-$1$ and type-$2$
fermions, we find that one set of the lattice is rigidly displaced
with respect to the other. Since each type of fermions has
opposite spin, the shifted lattices form an antiferromagnet-like
structure. It is called the spin density wave (SDW). We carry out
a Hartree-Fock computation for such a system, with the external
magnetic field tilted a variety of angles. The results show that
the cohesive energy of the shifted UCDW is steadily lower than
that of the shifted WC's, which means that the magneto-transport
anisotropy, similar to those for electrons in high LLs at half
fillings\cite{Lilly,Du1}, may take place in the lowest LL at
$\nu=2/5$ provided that the Zeeman energy is adjusted to an
appropriate range. The easy transport direction is always
perpendicular to the direction of the in-plane field. We also
calculate the dependence of cohesive energy on the partial filling
factor $\nu_1$. The result shows that the energy minimum appears
at $\nu_1=1/2$, which implies
that the half to half occupations of the $n=0\downarrow$-spin CF-LL and the $%
n=1\uparrow$-spin CF-LL is the most preferable state.

Suppose the two-dimensional electron system is confined by a harmonic
potential in the $z$-direction with the characteristic frequency $\Omega$.
The external magnetic field is tilted an angle $\theta$ with $\mathbf{B}%
=(B\tan\theta,0,B)$. According to G. Murthy \textit{et al}, the Hartree-Fock
theory in terms of CF variables gives a reasonably good account of physical
properties\cite{Murthy1}. The wave functions for the $n$th CF-LL are:
\begin{eqnarray}
\phi_{n,X}(\mathbf{r})&=&{\frac{1}{\sqrt{L_y}}}e^{-\frac{iXy}{l^2}}
\Phi_0^{\omega_+}((x-X)\sin\tilde{\theta}+z\cos\tilde{\theta})  \nonumber \\
& & \times \Phi_n^{\omega_-}((x-X)\cos\tilde{\theta}-z\sin\tilde{\theta}),
\label{one}
\end{eqnarray}
where $l$ is the magnetic length in the effective field and $X$ is an
integer multiple of $2\pi l^2/L$. $\Phi_n^\omega$ is the harmonic oscillator
wave function corresponding to the frequency $\omega$ and $\tan\tilde{\theta}%
=\frac{\omega_c^2}{\omega_+^2-\omega_c^2}\tan\theta$. The frequency $%
\omega_\pm$ are given by
\begin{equation}
\omega_\pm^2={\frac{1}{2}}(\Omega^2+\frac{\omega_c^2}{\cos^2 \theta}) \pm%
\sqrt{{\frac{1}{4}}(\Omega^2-\frac{\omega_c^2}{\cos^2 \theta})^2 +\Omega^2
\omega_c^2 \tan^2\theta}.
\end{equation}

The electron density operator is expressed in the momentum space as
\begin{equation}
\hat{\rho}(\vec{q})=\sum_{n,n^\prime,X}e^{-iq_xX} c_{n,X_-}^\dagger
c_{n^\prime,X_+}\rho_{n n^\prime}(\vec{q}),
\end{equation}
where $X_{\pm}=X\pm q_yl^2/2 $, $c_{n,X}$ destroys a CF in the state $%
\phi_{n,X}$. The matrix element $\rho_{n n^\prime}(\vec{q})$ can
be computed with the one-particle states Eq.(\ref{one}). The
Hamiltonian can now be written as\cite{Shankar}
\begin{eqnarray}
H&=&\frac{1}{2L_xL_y}\sum_{\vec{q},\{n_i\},\{X_i\}}v(q) e^{-iq_x(X_1-X_2)}
\rho_{n_1n_2}(\vec{q})  \nonumber \\
& &\times \rho_{n_3n_4}(-\vec{q})c^\dagger_{n_1,X_{1-}}
c_{n_2,X_{1+}}c^\dagger_{n_3,X_{2+}}c_{n_4,X_{2-}},  \label{two}
\end{eqnarray}
where $v(q)=2\pi e^2/q$. Equation (\ref{two}) is the correct form for the CF
Hamiltonian, and the energy coming from normal ordering represents the
Hartree interaction of an electron with its own correlation hole.

By using the standard manipulation, we can write explicitly the effective
potential $U_{nn^\prime}(\mathbf{q})$ as a sum of a Hartree term (in units
of $e^2/\kappa_0 l$)
\begin{equation}
H_{nn^\prime}(\mathbf{q})=\int\frac{dq_z}{\pi l}\frac{1}{q^2_\Vert+q^2_z} [%
F^\theta_{nn^\prime}(\mathbf{q})]^2
\end{equation}
and a Fock term
\begin{equation}
X_{nn^\prime}(\mathbf{q})=-2\pi l^2\int \frac{d\mathbf{p}}{(2\pi)^2}
H_{nn^\prime}(\mathbf{p})e^{i\mathbf{p}\times \mathbf{q}l^2},
\end{equation}
where $F^\theta_{nn^\prime}(\mathbf{q})$ is given by
\begin{eqnarray}
F^\theta_{nn^\prime}(\mathbf{q})&=&(n^\prime !/n!)^{1/2}
(\alpha^2/2)^{(n-n^\prime)/2}e^{-\gamma^2/4-\alpha^2/4}  \nonumber \\
& & \times L^{n-n^\prime}_{n^\prime} (\alpha^2/2)
\end{eqnarray}
with
\begin{eqnarray}
\alpha^2&=&(q_x\cos\tilde{\theta}-q_z\sin\tilde{\theta})^2 l_-^2
+q_y^2l^4/l_-^2\cos^2\tilde{\theta}  \nonumber \\
\gamma^2&=&(q_z\cos\tilde{\theta}+q_x\sin\tilde{\theta})^2 l_+^2
+q_y^2l^4/l_+^2\sin^2\tilde{\theta}.
\end{eqnarray}
Here $l_\pm^2=\hbar/m\omega_\pm$ and $L^m_n(x)$ is the Laguerre polynomial.

We adopt the Hartree-Fock (HF) approximation in a similar form introduced by
C{\^{o}}t{\'{e}} and MacDonald for double-layer systems\cite{Mac1,Mac2},
only the layer-index is replaced by the type-index in our case.

Allowing the charge density wave by making ansatz
\begin{equation}
<c^\dagger_{n X-Q_y l^2/2}c_{n^\prime X+Q_y l^2/2}> =e^{iQ_x
X}\Delta_{nn^\prime}(\mathbf{Q}),
\end{equation}
we carry out a HF computation on the UCDW and the triangular lattice. We
assume that $\Delta_{nn}$ are nonzero only for $n=0,1$\cite{Murthy}. The
cohesive energy can be calculated in the same way as it has been done for
the WC\cite{Yoshioka,Koulakov,Yang2}:
\begin{eqnarray}
E_{coh}&=&\frac{1}{2}\sum_{\mathbf{Q}\neq 0}\{ U_{00}(\mathbf{Q})+U_{11}(%
\mathbf{Q})  \nonumber \\
& & +2\cos (\mathbf{Q\cdot a})H_{01}(\mathbf{Q}) \} |\Delta_{00}(\mathbf{Q}%
)|^2,  \label{ground}
\end{eqnarray}
where $\mathbf{a}$ is the relative shift of the two sets of
lattice. Here we have not counted in the Zeeman energy and the
uniform direct interactions. The latter are cancelled with the
neutralizing positive backgrounds. The interactions of the
type-$1$ and type-$2$ fermions with the CFs in the fully filled
lowest $n=0\uparrow$-spin CF-LL subband are also omitted.

In our computations, the relative shift of the two sets of CDW reduces the
energy significantly. For the triangular lattice in the perpendicular
magnetic field, the cohesive energy of the unshifted lattices is -0.0961 (in
units of $e^2/\kappa_0 l$). The relative displacement ($\mathbf{a}=\{\frac{1%
}{2}\Lambda_b,\frac{\sqrt{3}}{6}\Lambda_b\}$ with lattice constant $\Lambda_b
$) reduces the energy to -0.1242. For the UCDW state, the energy reduction
for shifted lattice is even larger. This reduction of energy for shifted
lattice is attributed mainly to the lack of exchange symmetry between the
type-$1$ and type-$2$ fermions. Fig.1 shows that the shifted UCDW state is
always preferable to the shifted WC state for all of the tilted angles. As
is in the case of high LLs, a stable UCDW ground state or stripes, leads to
the anisotropy in magneto-transport experiment. As each set of the UCDW has
opposite spin polarization, the shifted lattice forms a pattern of
antiferromagnet. We call this new phase the unidirectional spin density wave
(USDW). The space period of the USDW is $\sim 5.2 l$. For completeness, we
have also considered the triangular lattice of "bubbles", with each bubble
containing in general several electrons by using the scheme developed by
M.M. Fogler \textit{et al}\cite{Fogler}. Our result shows that the lattice
of one electron per "bubble" has the lowest cohesive energy.

In Fig.2 we calculate the cohesive energy of shifted UCDW by varying the
partial filling factor $\nu_1$ from $0$ to $1$ for zero in-plane magnetic
field. We assume that the space periods of the CDWs are the same for both
CF-LLs bands. It can be seen that the energy minimum appears at half
filling, which means $\nu_1=\nu_2$. This figure reflects the typical
characteristics of Fig.1 in Ref.\onlinecite{Apalkov}. It indicates that the
state of half to half occupation for $n=0\downarrow$-spin CF-LL and the $%
n=1\uparrow$-spin CF-LL is energetically more stable than the full
occupation of only one sub-band of the CF-LLs. Taking account of the fully
occupied $n=0\uparrow$-spin band, the total polarization is one-half.

Because of the small $g$ factor of electrons in $GaAs$, spins may not be
fully polarized in FQH states. Transitions between singlet, partially
polarized, and fully polarized states for a number of fractional fillings
can be understood in terms of CFs with a spin\cite{Jain1,Du2,Khan}. For $%
\nu=2/5$, the transition takes place when the unoccupied $n=1 \uparrow$-spin
CF-LL subband crosses the occupied $n=0\downarrow$-spin CF-LL subband as the
Zeeman energy increases. If the cross is trivial, nothing interesting will
happen. However, the competition between exchange and direct Coulomb
interaction of CFs results in the spontaneously breakdown of translational
symmetry.

The mechanism of the USDW in this work is to some extend similar to the
isospin stripes at integer filling factors for the double-layer systems at $%
\nu=4n+1$\cite{Brey}, in which the real spin-index is replaced by the
isospin-index. The origin of such isospin stripe order is a competition
between the exchange and the direct Coulomb interaction. Exchange favors
accumulating all the electrons in one layer(in order to maximize the isospin
exchange field), whereas direct Coulomb energy is lower when the electron
density is distributed uniformly between the layers. It is also interesting
to compare our prediction of USDW to the anisotropic transport for
even-number filling factors ($\nu=4,6,\cdots$) when the magnetic is tilted
to a large angle\cite{Pan}. It can easily be seen that the unidirectional
spin density wave in this work is just a CF-version of the unidirectional
spin density wave of electrons suggested in Refs.\onlinecite{Pan, Demler}.

Hitherto, transport anisotropy was experimentally observed only in high LLs%
\cite{Lilly,Du1}. Our results predict that in the lowest LL, there may also
exist stripe phases at some fractional filling factors ($\nu=2/5, 3/7,\cdots$%
), provided that the Zeeman energy $E_Z$ is set to an appropriate
value. Since the space period of the USDW is large enough, the
USDW is stable against the quantum fluctuations in the lowest LL.
We note that S.Y. Lee \textit{et al}\cite{Lee} proposed another
kind of spontaneous stripe order at certain even-denominator
fractions in the lowest LL. They argued that for Landau level
filling factor of the form $\nu=(2n+1)/(4n+4)$, which corresponds
to CF filling factor $\nu^*=n+1/2$, the system phase separates
into stripes of $n$ and $n+1$ filled CF-LLs. Our picture of USDW
essentially differs to theirs.

In summary, we have computed the ground-state energies of USDW and WC
consisting of composite fermions. We find that the relative shift of the two
sets of interacting CDW lattices reduces remarkably the cohesive energy of
the partially polarized density wave state of CFs proposed by G. Murthy\cite
{Murthy1}. As the external magnetic field is tilted an angle, the USDW is
always energetically preferable, which means that anisotropic transport may
be observed in the lowest LL at $\nu=2/5$ if the Zeeman energy is properly
adjusted. We also found that half to half occupation of the $n=0\downarrow$%
-spin CF-LL and the $n=1\uparrow$-spin CF-LL is more stable than the full
one-band occupation. Experimental observation of the stripes in the lowest
LL will support the concept of such antiferromagnet-like charge/spin density
wave of composite fermions as we have described in the context.

This work was supported in part by NSF of China. S. J. Yang was supported by
China Postdoctoral Science Foundation.

\begin{center}
FIGURES
\end{center}

Figure 1 The cohesive energies (in units of $e^2/\kappa_0 l$) of shifted
USDW and shifted WC versus tilted angle $\theta$ for $\Omega/\omega_c=3.0$.
The inset is the rescaled energy for USDW. It can be seen that the cohesive
energy of the USDW state decreases as the tilted angle increases.

Figure 2 The cohesive energy $E_{coh}$ of USDW versus partial filling factor
$\nu_1$. The minimum of energy at $\nu_1=1/2$ indicates that the half to
half occupation of CF spin-subbands is energetically preferable to the full
occupations of only one CF spin-subband.

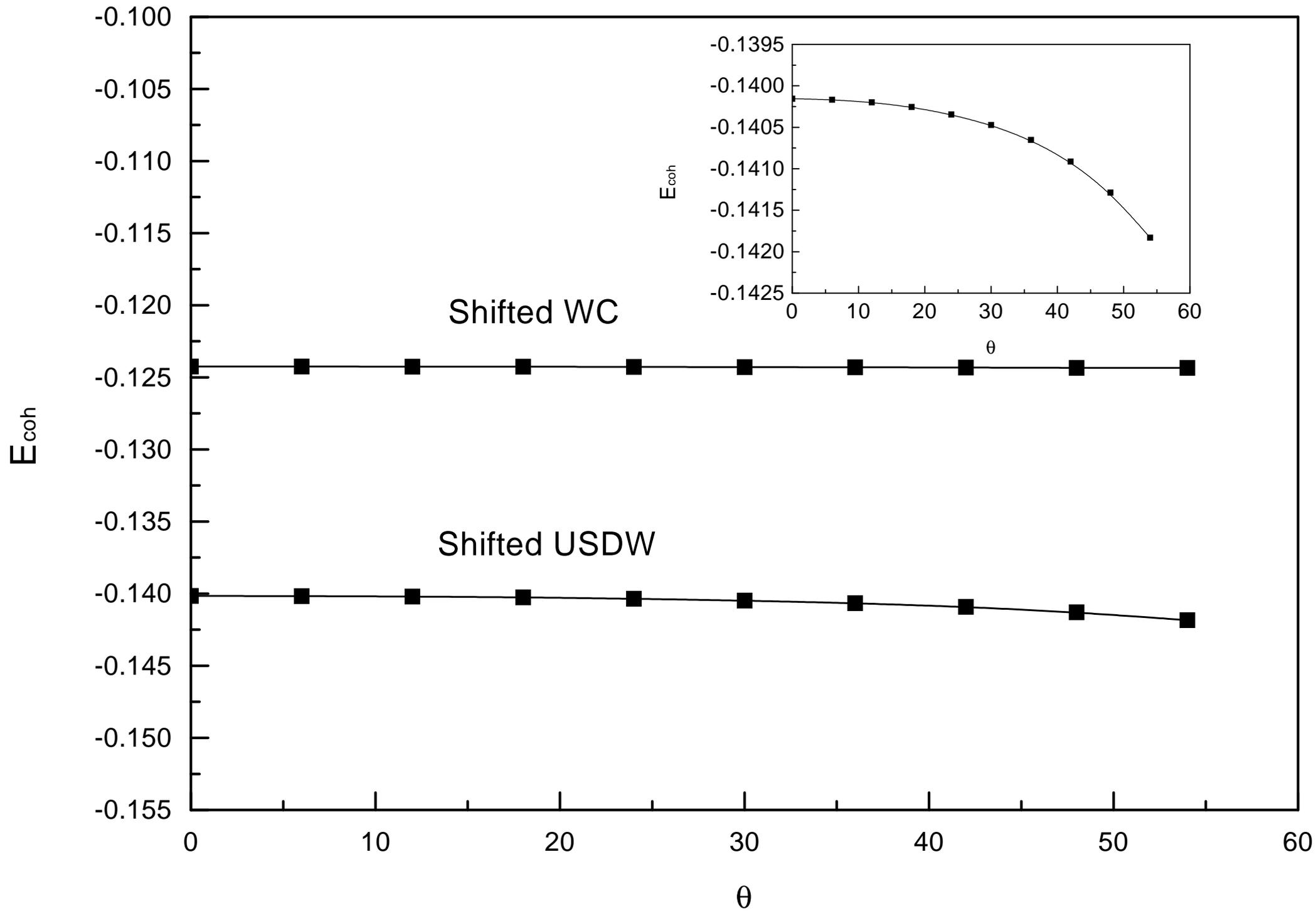
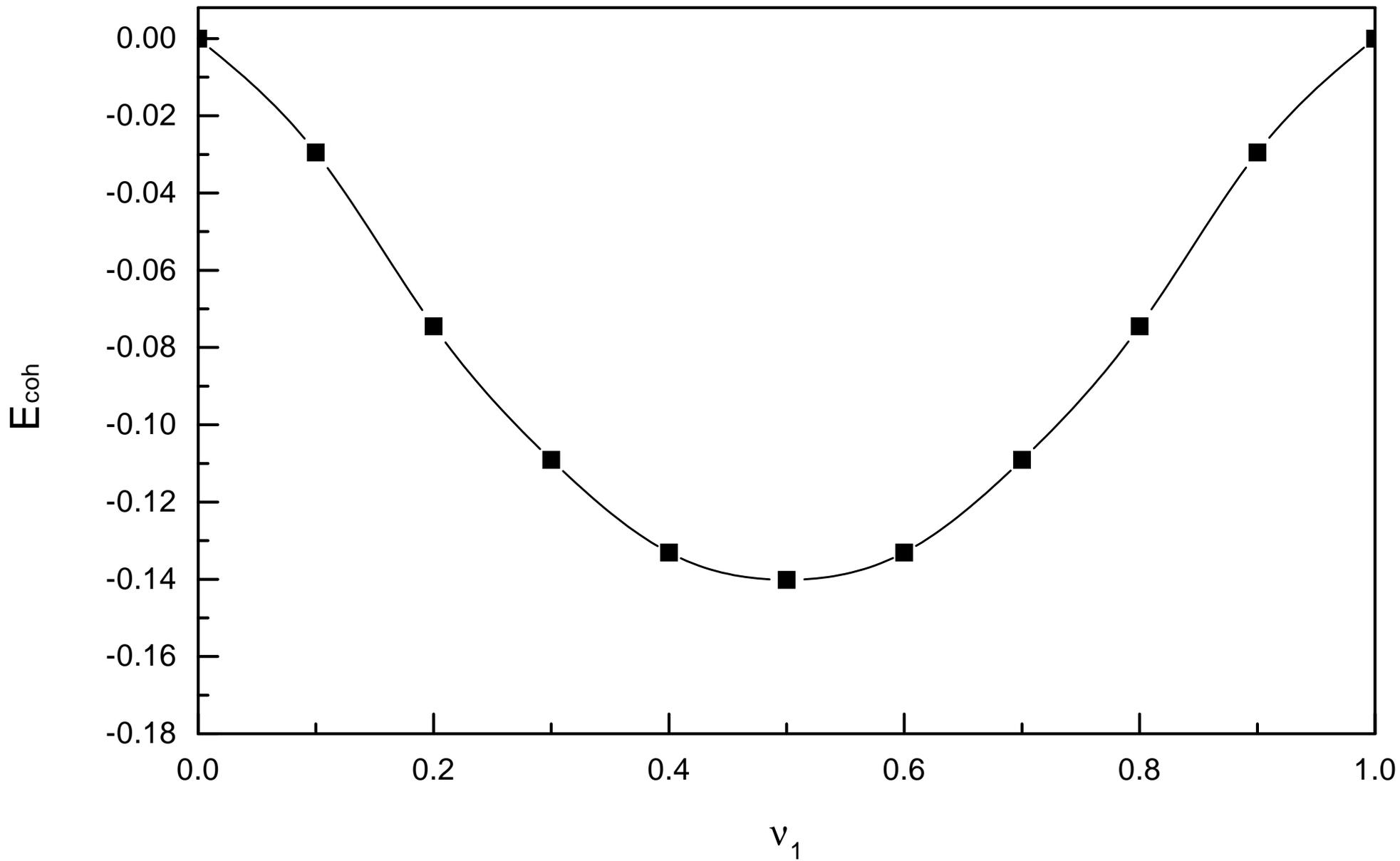
\end{document}